\documentstyle[prb,aps,multicol,psfig]{revtex}
\begin{document}
\title{
Surface reconstruction induced geometries of Si clusters
}
\author{ Efthimios Kaxiras$^{\dagger}$ \\
Institute for Theoretical Physics\\
University of California, Santa Barbara, CA 93106
}
\maketitle
\begin{abstract}
We discuss a generalization of the surface reconstruction arguments
for the  structure of intermediate size Si clusters, which
leads to model geometries for the sizes 33, 39 (two isomers), 45 (two 
isomers), 49 (two isomers), 57 and 61 (two isomers).
The common feature in all these models is a structure that closely 
resembles the most stable reconstruction of Si surfaces, 
surrounding a core of bulk-like tetrahedrally bonded atoms.
We investigate the energetics and the electronic structure of
these models through first-principles density functional theory
calculations.  These models may be useful in understanding
experimental results on the reactivity of Si clusters and their shape
as inferred from mobility measurements.

\end{abstract}

\begin{multicols}{2}

\section{Introduction}
\label{intro}

Silicon is a material that attracts considerable interest
due to its technological importance.  It has also come to be regarded as 
the representative covalently bonded solid.
In the last decade, a new form of Si has generated much 
excitement because it holds promise for 
novel technological applications as well as for providing 
unique insight into the nature of covalently bonded materials \cite{Brus}.
This new form is clusters of Si, which consist of a few tens to 
a few hundreds of atoms and have properties 
different from the bulk.
The properties of Si clusters
depend on their size, and a detailed study of this
dependence could potentially lead to a 
better understanding of how chemical bonding evolves 
from that characteristic of small molecules to that characteristic
of the bulk.  

Despite much theoretical and experimental work on Si clusters
during the last decade, little is known about their structure.
It is self evident that for very small cluster sizes, all of the atoms 
in a cluster will be exterior atoms, in the sense that there is 
no shell formed by a subset
of atoms that completely surrounds any one atom.
These very small Si clusters should therefore be viewed 
as molecules, i.e. entities with unique structural, physical
and chemical characteristics that do not resemble 
other forms of Si.  
This picture was originally proposed by Phillips
\cite{Phillips10}, who analyzed early experiments on 
fragmentation patterns
\cite{Bloomfield,Smalley0}.
Equally evident is the 
fact that beyond a certain size there will be interior atoms
completely surrounded by a shell of other exterior atoms. 
(In the following, we use the terms ``interior'' and ``exterior'' 
as defined above to 
characterize cluster atoms, and reserve the terms 
``bulk'' and ``surface'' to characterize atoms of the solid.)
It is also natural to expect that for a large enough size, the
cluster will resemble a crystalline fragment.  

The size at which the transition from molecules to 
bulk-like fragments takes place
has not yet been determined precisely, although 
several theoretical predictions have been made
\cite{Cheli1,Tomanek}.
It has also been proposed
that the experimentally 
observed sharp transition in the shape of Si clusters \cite{Jarrold}, 
which occurs at around size 27 
may be related to the onset of
structures with interior (``bulk''-like) and exterior (``surface''-like)
atoms in the cluster \cite{KJ}. 
Such a transition in shape had actually been predicted by theoretical  
simulations \cite{Cheli2}.  

The structure of the small Si clusters (up to 10 atoms) has been 
determined by extensive theoretical calculations
\cite{Raghavachari,Ballone,Binggeli} 
and by comparison of calculations to spectroscopic measurements
\cite{Honea,Raghavachari-Si-small,Li-Si-ir,Cheli-vibr,Cheli-photo}.  
Much less is known for the larger clusters, although a number 
of interesting models have been proposed.
One possibility is models analogous to fullerene cages
\cite{Nagase,Menon-Si60}, but 
given the tendency of Si to form four-fold coordinated structures
this seems unlikely.
A different approach is to consider models 
with non-spherical shapes for medium sizes
\cite{KJ,Grossman-long,Rama-growth} 
or with interior and exterior atoms for larger sizes, but 
\cite{Kax1,Patterson,Rama}.
The assumption that there exist interior and exterior atoms does not
necessarily imply that their respective environment will 
resemble that of bulk or surface atoms.
Nevertheless, it is an attractive proposition to consider
what types of structures can be produced if geometries 
resembling the bulk structure (for interior cluster atoms) and 
the surface reconstructions (for exterior cluster atoms)  
were to dominate the cluster geometry.  
Only a few cluster sizes are compatible with  
the requirement that {\em all interior cluster atoms}
have environments that closely resemble the bulk structure, and 
{\em all exterior cluster atoms} have environments that
closely resemble surface reconstructions.
We call the resulting models Surface Reconstruction
Induced Geometries (SRIG's for short).
This idea has been invoked as an explanation of the 
existence of magic numbers in Si clusters \cite{Kax1} as revealed 
by their reactivity with various chemical agents 
\cite{Smalley1,react1,react2,react3,react4,Smalley2}.

In this paper, we present 
in Section \ref{srig} a thorough investigation of SRIG models 
for the structure of selected  
clusters spanning the sizes from 33 to 61 atoms.  
In Section \ref{properties} 
we use first-principles density functional theory calculations
to study the energetics and electronic properties of these 
models.  Finally, in Section \ref{discussion} we comment on the relevance 
of these models to experimental observations
of the reactivity and the mobility of Si clusters.

\section{Surface reconstruction induced geometries}
\label{srig}

Based on the assumption that the surface reconstruction determines
the cluster geometry, 
we have constructed models 
for Si clusters in which exterior atoms resemble closely the
coordination and bonding of atoms in the most stable surface
reconstructions of Si, including the (111) $7 \times 7$ 
dimer-adatom-stacking fault (DAS) reconstruction \cite{Takayanagi}, 
the (100) $2 \times 1$ dimer reconstruction, and the 
(111) $2 \times 1$ $\pi$-bonded chain reconstruction \cite{Pandey}.
All interior atoms have the 
coordination and bonding arrangement of bulk Si atoms, i.e.
four-fold tetrahedral coordination.  The special sizes for 
which we were able to construct such models include 
the sizes 33, 39, 45, 49, 57 and 61.
This does not exclude the possibility of other sizes
that have similar bonding arrangements in this size range. 
In order to construct these models we imposed two additional 
constraints: (i) that the clusters have an atom at their geometric
center, and (ii) that the clusters have tetrahedral overall symmetry.
These constraints were motivated by simplicity considerations 
and not from any physical requirements.  If the constraints 
are relaxed, it might be possible to construct  
additional structures with similar characteristics
but lower symmetry.  

In describing the cluster geometries, we adopt the following conventions:
We will display {\em only the exterior atoms} of the cluster,
in a two-dimensional picture which is obtained by unfolding three
of the 
four sides of the tetrahedron symmetrically around the
fourth side.  In so doing, one
obtains an equilateral triangle, at the center of which there
is a smaller, inverted equilateral triangle corresponding to one side of 
the tetrahedron around which the other three sides have been 
unfolded.  We denote the sides of the tetrahedron by 
dotted lines in all the structural figures.  
In the models we discuss there exist one to five interior atoms,
which are not shown in the structural figures.  The position 
of these interior atoms is discussed in detail along with 
every structural model.  
What is actually shown in the structural figures are the projections
of atoms on the faces of the tetrahedron.  For simplicity we refer
to these projections as the atomic positions. 
This manner of displaying the cluster 
geometries has the added advantage that the reader can 
easily construct schematic three-dimensional models 
of the cluster exterior by cutting out the 
two-dimensional figures, folding them at the three edges of the 
central inverted triangle, 
and gluing the outer edges together.

We have also adopted a shading scheme which is common to all the 
cluster sizes:  First, we indicate by open 
circles (we call those white atoms) the exterior atoms that are at the 
apexes of the tetrahedron; there are four such 
atoms in each model. 
In some of the models (45B, 61A, 61B) 
there exists an additional set of four
atoms at the centers of the faces of the 
tetrahedron, which we also denote by open circles.
Next, we indicate 
by hatched circles the atoms that 
are bonded to the first set of white atoms; there are twelve 
such atoms in each model.
Next, there are atoms that are bonded to the twelve hatched
atoms, and are indicated by grey circles.  
Finally, there are atoms shown in black,  
which are bonded to the grey   atoms or to the white 
face-center atoms, but not 
to the hatched or white apex atoms. 
Sets of atoms with the same shading are equivalent by symmetry.
With these conventions, we proceed to describe the individual 
cluster models in detail. 

We begin with the cluster of size 33,
shown in Fig. 1.  This cluster had been proposed
previously \cite{Kax1}, and consists of 28 exterior atoms and 5 interior atoms.
Of the five interior atoms, one is at the geometric center of the 
tetrahedron and the other four are bonded to the central atom 
in directions pointing away from the center toward the vertices 
of the tetrahedron.  All interior atoms are four-fold coordinated
as atoms in bulk Si are.  The neighbors of the four interior
atoms other than the central one are the twelve hatched exterior atoms.
The hatched exterior atoms are also four-fold coordinated, but 
their bonds are at angles severely distorted relative to the 
tetrahedral angle.  The neighbors of the hatched exterior
atoms are the four interior atoms (other than the central one)
and the twelve grey  exterior atoms
which are three-fold coordinated: they have two bonds to the 
hatched exterior atoms and one bond to another grey exterior 
atom each.  Finally, there are four exterior white atoms
at the apexes of the tetrahedron;  these atoms are three-fold 
coordinated, and are bonded to three hatched exterior atoms.

The close similarity of this structure with surface reconstructions
of Si involves two elements:  (a) The four white apex atoms are in local
geometries which are very similar to the adatoms on the 
Si(111) $7 \times 7$ reconstruction: the adatoms are directly 
above second layer bulk Si atoms (in the present model this role is played by
the four interior atoms other than the central one).
(b) The twelve 
grey exterior atoms are bonded in pairs like the dimers on the 
Si(100) $2 \times 1$ reconstruction;  these atoms also play the 
role of the dimers that characterize the Si(111) $7 \times 7$
reconstruction and help stabilize the adatom reconstruction.
Thus, in this model there is a cooperative effect that combines features
from the two most stable surface reconstructions of Si
to produce a highly symmetric structure with 5 interior and 28
exterior atoms.  

It is noteworthy that the bonding 
arrangement of exterior atoms in this structure
consists of pentagonal rings (with corners at one white, 
two hatched and two grey atoms) and distorted hexagonal rings
(they appear to be almost triangles in Fig. 1, with corners at
three hatched and three grey atoms).  In this sense, the shell
of exterior atoms resembles the structure of the C$_{28}$
fullerene, which has been studied extensively as the smallest
stable fullerene \cite{C28-str,C28-exp,C28-the}.  
The resemblance, however, is superficial, since
it is limited to this 
geometric aspect, while the C$_{28}$ fullerene is stabilized
by the $\pi$-bonding of atoms at the six hexagonal rings and 
the Si$_{33}$ structure is stabilized by the presence 
of the adatom and dimer surface features, as well as the 
presence of the five interior atoms.  Additional important differences
concern the electronic properties of the two models, with 
the C$_{28}$ cluster being an open shell structure
\cite{C28-the}, while the 
Si$_{33}$ model is a closed shell structure (see Section \ref{discussion}).

We next consider the two models for the Si cluster with 39 atoms.
The first model (referred to as 39A,
shown in Fig. 2) is similar to the 33 atom model discussed
above, with six additional atoms that form bonds to the grey
atoms (shown as black, at the midpoints of tetrahedron
edges).  In this model, the white, hatched and 
grey atoms are coordinated as in the 33 atom
model.  The six extra atoms are placed between 
pairs of grey atoms, and form new bonds to those atoms 
which become four-fold coordinated.  The six extra atoms 
are two-fold coordinated each.  This type of atomic arrangement 
is not usually observed on Si surfaces.  However, it is a reasonable 
arrangement for Si atoms which cannot find close neighbors to bond to.
The preferred bonding for these two-fold coordinated 
atoms will be through $p$ orbitals
to their two neighbors, while they
retain two of their four valence electrons in a low-energy
non-bonding $s$ orbital. 

The second model for the 39 atom cluster 
(referred to as 39B, shown in Fig. 3), consists of a different 
geometric pattern. The four white apex atoms are in
similar arrangements as in the 39A model. 
The hatched atoms have three bonds, one to interior atoms,
one to white apex atoms and one to black atoms at the
centers of the tetrahedron edges. 
The grey atoms no longer form dimers;
instead, they form trimers centered at the faces of the tetrahedron.
This is a somewhat unusual arrangement for Si atoms, and 
is not encountered in native surface reconstructions of Si, since
the presence of trimer units induces significant
strain on the surface. 
However, a trimer
reconstruction of this type is common for group-V atoms 
on the surfaces of III-V semiconductors
\cite{III-V}, or when the group-V elements
are used as passivating layers on the Si(111) surface
\cite{Si111-V}.  Moreover, group-IV
elements like Sn and Pb, also form trimer
units on the Si(111) surface \cite{Si111-IV}.  
We suggest that under the proper conditions (represented
here by the size of the cluster) similar trimer units of Si atoms
may be stable on the cluster surface.   Finally, there are six
atoms at the centers of the edges of the tetrahedron, which in this model
are four-fold coordinated with two bonds to the hatched atoms
and two more bonds to grey atoms in the trimers.   This arrangement 
results in three-fold and seven-fold rings on the exterior shell of the
cluster and is markedly different from the previous geometry for
the size 39.

The first model we considered for the 45 atom cluster 
(referred to as 45A, shown in Fig. 4) is a simple
variation of the 39B model: the trimers at the centers of the 
tetrahedron faces are replaced by hexagons and the 
black atoms at the centers of the 
tetrahedron edges are eliminated.  This should help reduce
the strain induced by the trimers, and it should enhance the
stability of the apex atoms: the coordination of their neighboring
hatched atoms has been restored to four (one apex atom, one interior atom
and two grey atoms), while the exterior pattern again 
contains three
pentagonal rings around each white apex atom and dimers of grey atoms.
In addition to the pentagonal rings
(with corners at one white apex, two hatched and 
two grey atoms), 
the exterior of the 45A model contains
two types of hexagons: one composed of 
six grey atoms and one composed of four grey and two hatched
atoms.  In this sense, the exterior of this model bears 
a superficial resemblance to the C$_{40}$ fullerene structure \cite{Rama}, 
but just like in 
the case of Si$_{33}$ and its companion fullerene C$_{28}$, 
the stability and electronic properties of the two types of 
clusters are quite different.   

The next model for a 45 atom Si cluster (45B) is illustrated in
Fig. 5, and corresponds to a 
different type of exterior bonding arrangement \cite{Kax1}.
In this case there is only one interior atom, bonded to the 
four white atoms that occupy the centers of the tetrahedron faces.
The white atoms at the tetrahedron apexes 
are no longer at positions that resemble the adatom structure 
of the Si(111) $7 \times 7$ reconstruction.
Instead, they form intersection points where 
zig-zag chains consisting of hatched and grey atoms meet.  The
presence of these chains should produce a low energy exterior shell
because they resemble closely the chains of atoms in the other stable
reconstruction of the Si(111) surface, the $\pi$-bonded
chain model proposed by Pandey \cite{Pandey}.
In this model, the dangling bonds on the neighboring
chain atoms form $\pi$-bonded combinations.
The geometry of the exterior shell consists of pentagonal rings around the
four white apex atoms and hexagonal rings
around the four white face-center atoms.
Since there is only one interior atom bonded to the
four face-center white atoms, the hatched atoms are only
three-fold coordinated, with all their neighbors belonging to the
exterior shell.  The grey and black atoms are three-fold coordinated,
as are the apex white atoms.  The only atoms that are four-fold
coordinated, other than the central atom, are the four white atoms
at the centers of the tetrahedron faces.
This model has chirality
(i.e. there exist equivalent left and right handed versions), which 
is evident from the fact that the zig-zag chains do no lie on any
high symmetry direction of the tetrahedron, and thus break the 
left-right symmetry of the previous models.

The same exterior shell can be used to construct two different
models for the 49 atom cluster.  We refer to these models as 49A and
49B.  The first (49A) constists of the exterior
shell shown in Fig. 5 plus five interior atoms, one at the
geometric center of the
tetrahedron and four more bonded to the central atom, and pointing
away from it toward the four white apex atoms.  The second (49B) consists
of the same exterior shell and five interior atoms again, one at the 
geometric center of the tetrahedron and four more bonded to the
central one and pointing away from it, toward the four white
atoms at the centers of the tetrahedron faces.
In the case of model 49A, the four white apex atoms have a coordination
similar to that
of adatoms on the Si(111) $7 \times 7$ DAS reconstruction,
as described in the case of the 33 atom model.
In the case of model 49B, the role of adatoms is assumed by the four
white atoms at the centers of the tetrahedron faces.  In both cases,
the atoms that are bonded to the adatoms become four-fold coordinated.
Specifically, in model 49A the hatched atoms become four-fold coordinated,
while in model 49B the black atoms 
become four-fold coordinated.
Since the exterior geometry in models 49A and 49B 
is the same as in the model 45B,
the pattern of pentagons and hexagons is the same.
The important difference between models 49A and 49B 
is that in the first the adatoms
(white apex atoms) are surrounded by three pentagonal rings, while in
the second the adatoms (white face-center atoms) are surrounded by 
three hexagonal rings.

The model for the 57 atom cluster is an extension of the 
45A model, in which the six-fold rings of grey atoms 
on the tetrahedron faces are replaced by nine-fold rings
as shown in Fig. 6.
This is achieved by the addition of three
extra atoms (shown in black) on each tetrahedron face.
The rest of the cluster structure is the same
as the 45A model, with five interior atoms (one at the geometric
center of the tetrahedron and four more bonded to it, pointing toward
the four white apex atoms) and the four white apex and twelve hatched atoms
in similar positions as before.  The extra atoms that turn the 
face-centered hexagonal rings of the 45A model 
into nine-fold rings, are themselves
bonded in dimer-like pairs across the the centers of the
tetrahedron edges.  In this way the exterior shell is composed
of nine-fold rings centered at the tetrahedron faces and two 
types of pentagonal rings, one with corners at a white apex,
two hatched and two grey atoms, and one with corners at 
a hatched, two grey and two black atoms.

The two models we have constructed for the 61 atom cluster
involve geometries that have four white face-center atoms, in 
two different configurations.
The first configuration (labeled 61A, shown in Fig. 7)
derives from the 57 atom model, with
the additional white atoms bonded to the three black atoms on
each face, and with the dimer bonds between black atoms broken. 
The remaining atoms are in exactly the same configuration as in the
57 atom model.  This change modifies the exterior shell,
which now consists of eight-fold rings (centered at the
edges of the tetrahedron and composed of two hatched, four grey and
two black atoms) and two types of pentagonal rings, one composed of
a white apex, two hatched and two grey atoms, and the other
composed of a white face-center, two black and two grey atoms.
In this model, as far as the exterior shell is concerned, the
white apex atoms and the white face-center atoms are equivalent.
Therefore the four interior atoms (other than the central one)
can be considered to be pointing toward either the white apex or the 
white face-center atoms.

Finally, the second model for a 61 atom cluster
(labeled 61B, shown in Fig. 8) contains
four white apex atoms at adatom positions, surrounded by
hatched atoms that form pentagonal rings with adjacent grey atoms,
and four white face-center atoms that form hexagonal rings 
with a hatched, two grey and two 
black atoms.  In this case the four white apex atoms
and the four white face-center atoms are not equivalent.  The
four interior atoms (other than the central one) are bonded to
the four white apex atoms, since those have the benefit of
being surrounded by pentagonal rings which should stabilize their
adatom-like features.
In the 61B model, the exterior shell consists of pentagonal
and hexagonal rings, which gives it a 
superficial resemblance to the corresponding
C$_{56}$ fullerene structure.

\section{Structural, energetic and electronic features of SRIG models}
\label{properties}

Having described the geometric features of the SRIG models,
we discuss next the results of energy optimization and electronic
structure calculations for these models.
The calculations are based on density functional theory in the
local density approximation \cite{DFT-LDA} (DFT-LDA.  
Although these calculations do not
necessarily provide the most accurate comparisons of cluster 
energies, they are reasonably reliable in determining optimal
geometries by minimizing the magnitude of the 
calculated Hellmann-Feynman forces.  The present calculations employ
a plane wave basis with a cutoff energy of 8 Ry, and a
cubic supercell with lattice constant equal to 15.875 \AA
(which gives a basis of 10400 plane waves), 
for the clusters of sizes 33 to 49, and equal to 17.463 \AA, for 
the clusters of sizes 57 and 61
(which gives a basis of 13600 plane waves).  With these supercells the 
periodic images of the clusters are reasonably well separated.
A single k-point (the center of the supercell Brillouin Zone)
was used in all the cluster calculations. 
The Car-Parrinello iterative scheme \cite{CP} was used to solve the Kohn-Sham
equations self consistently, and non-local norm conserving pseudopotentials
were employed to model the atomic cores \cite{Bachelet}.

In Table I we provide a list of the atomic positions for
the SRIG models.  In this Table only the positions of inequivalent atoms
are given, with the remaining positions obtained by applying 
symmetry operations of the tetrahedral group.
The position of the central atom in each cluster is taken 
to be the origin of the coordinate system and is omitted from the
list of positions.

We discuss first certain structural features of the optimized cluster
geometries.
These considerations were motivated by some exciting recent experiments
that are able to resolve aspects of the cluster
geometry through gas phase mobility measurements \cite{Jarrold-mobil}.
In the simplest picture, the mobility measurements reveal the 
projected cross section of the cluster. 
For spherically shaped clusters, this cross section is given 
by $\Omega = \pi b_{min}^2$, where $b_{min}$ is the distance 
of closest approach between gas phase molecules and the cluster.
By considering the atoms on the surface cluster as points, we 
can obtain $b_{min}$ as the distance of the atom farthest from 
the cluster center, which gives the values of $\Omega$
tabulated in Table II.  This quantity is
a coarse measure of the cluster geometry and gives approximately the 
overall cluster size in cross section.  
From the comparison of $\Omega$ values in Table II, it is evident 
that the cluster cross section does not follow exactly the number of  
atoms in the cluster.  It is also interesting that the clusters with 
the highest cohesive energy (see following paragraph and Table II)
also have the smallest $\Omega$
(especially the models 33, 45A, 49A and 49B), i.e. 
the lowest energy clusters are also 
the most compact ones.

As Shvartsburg and Jarrold
have pointed out \cite{Jarrold-convx}, the projection of the cluster
cross section is appropriate for clusters which are locally convex.
When the surface of the cluster contains concave regions, the 
scattering cross section is increased due to multiple collisions.
In order to obtain a measure of this effect, we have calculated the
ratio $R_c$ of surface triangles that are concave to the total 
number of triangles on the cluster surface.
Since this is a measure of local concavity, we have limited 
the definition of surface triangles to those which have sides 
smaller than 6.35 \AA.  Table II gives the values of $R_c$ for the 
various models.  Two of the clusters, 33 and 45A, are 
everywhere locally convex ($R_c = 0$).
For the rest of the clusters, $R_c$ ranges between 1.9\% to 10.8\% 
which is relatively small, but not insignificant, since the 
accuracy of mobility measurements \cite{Jarrold-convx}
is about 2\%.
The clusters with the smallest non-zero $R_c$ are 49A and 49B,
which together with the clusters 33 and 45A form the group
of clusters which are most compact (lowest $\Omega$) and 
have the highest cohesive energy. 
These comparisons should be helpful in identifying which of the 
SRIG models may be relevant to the structure of Si clusters as 
determined by mobility experiments. 

For the energetic comparisons, 
since the type of calculations employed here
do not provide an accurate estimate
of the cohesive energy of bulk Si, we have opted to quote the cohesive
energy per atom of each cluster by comparing the cluster
energy to an equivalent number of bulk Si atoms, and
using the experimental value for the cohesive energy of the
bulk (4.68 eV).  In this way, the quoted energies per atom for the clusters
should be closer to realistic cohesive energies.
For the bulk calculation, we use the same plane wave cutoff,
the primitive 2-atom unit cell of the diamond lattice, and 
a set of k-points that produces a density in reciprocal 
space 
equivalent to the center of the supercell cube in the cluster calculations.

Table II contains a comparison of the cohesive energy per atom
and the corresponding energy gap between the highest occupied
molecular orbital (HOMO) and lowest unoccupied molecular orbital (LUMO),
for the different clusters we studied.
The model with the highest cohesive energy per atom (49A)
has almost 82\% of the bulk cohesive energy, while the model
with the lowest cohesive energy (39B) has 77\% of the bulk cohesive
energy.
Fig. 9 shows the density of states (DOS) for each model.
The symmetry of the clusters dictates that the electronic
states are singly, doubly or triply degenerate, according to the
dimensions of the irreducible representations of the tetrahedral group.
In all cases the total band width (between 12 and 13 eV)
is comparable to that of the bulk.

There are some interesting insights revealed by these comparisons
of cohesive energies and electronic structure.
The cluster with the smallest cohesive energy (least 
stable energetically) is the cluster with 39 atoms and trimer units
at the tetrahedron surfaces (model 39B, Fig. 3).
The relatively high energy of this cluster can be rationalized
as being due to the strain induced by the trimer units.
Interestingly, this cluster also has the largest HOMO-LUMO gap.
Thus, at least for the structures considered here, there is no
direct correlation between the energetic stability and the 
HOMO-LUMO gap.  The existence of a large HOMO-LUMO
gap may be indicative of low chemical reactivity, since in this case
the system has a closed electronic shell.  In this sense, if the 39B 
SRIG model were to be formed, it might be expected to have low chemical
reactivity.  However, a quantitative study of chemical reactivity
should involve a more detailed examination of cluster electronic
states, as well as case studies of chemical reactions between
the cluster and representative molecular agents \cite{Rama-react}.
The presence of a different model of the same size (39A)
which has lower energy and a comparable HOMO-LUMO gap suggests that
the likelihood of the geometry 39B being realized is small.

Another interesting point is that the lowest energy clusters
are the two models with 49 atoms, closely followed by the 45A model
and the 33 model.
While the 45A and 33 models have significant HOMO-LUMO gaps,
the two 49-atom models have very small (49A) or non-existent (49B) gaps.
Thus, based on the simple arguments mentioned above on the 
relation of the HOMO-LUMO gap to chemical reactivity, one
might expect that the models 33 and 45A will exhibit low 
chemical reactivity, while models 49A and 49B will have higher
chemical reactivity, despite their lower energy per atom.
This is consistent with the experimental findings of Jarrold and Honea
\cite{Jarrold-anneal}, who observed that chemical reactivity
and thermodynamic stability are not related. 
Finally, the larger clusters of sizes 57 and 61 again have
very small (57) or non-existent (61A, 61B) gaps,
suggesting high chemical reactivity despite their relative 
energetic stability (especially for model 61A).

\section{Discussion and conclusions}
\label{discussion}

We have presented a detailed discussion of the geometric features,
the relative energies and the electronic structure of 
models for Si clusters of sizes 33, 39, 45, 57 and 61.
The basic characteristic of these models is their resemblance
to bulk Si, including geometries of the exterior atoms
that resemble surface reconstruction features and geometries
of the interior atoms that resemble the four-fold tetrahedral
coordination of the bulk.  Although this
is an appealing feature, it does not guarantee that
these models are the energetically preferred geometries.
In fact, it seems that molecular dynamics (MD) or
simulated annealing (SA) simulations based on empirical \cite{Jelski},
semi-empirical \cite{Menon-Si45} 
and first-principles calculations \cite{Roth,Mukher}
tend to give geometries different from the ones proposed here.
Therefore, it is important to address to what extend 
the models considered here are relevant to the 
structure of real Si clusters. 

We suggest that these models may indeed be relevant to the
structure of real Si clusters for the following reasons:

(1) It appears from experiment that there is something
special about the chemical reactivity of the (so called ``magic'')
sizes 33,
39 and 45, while other clusters have approximately constant
and much higher (by several orders of magnitude) reactivity 
from these magic numbers \cite{Smalley1,react1,react2,react3,Smalley2}.  
It is appealing
that for all the magic number sizes we were able to construct SRIG models,
and that these particular models have the largest HOMO-LUMO gaps of all
the structures considered.
Moreover, 
the SRIG models of sizes 49, 57 and 61, which are 
energetically
equally stable stable as the magic numbers,  
have very small or non-existent HOMO-LUMO gaps.
In this sense, the pattern of
SRIG models is compatible with the pattern of experimental
measurements of reactivity, assuming that the HOMO-LUMO gap
can be used as a coarse measure of 
chemical reactivity (with the caveats mentioned in Section 
\ref{properties}).  

(2) If the clusters of various sizes possess no special
geometric features, as the simulations based on various
methodologies suggest \cite{Jelski,Menon-Si45,Roth,Mukher}, 
then it is difficult to explain
the dramatic changes in reactivity and the existence of the magic numbers.
Actually, MD simulations based on DFT-LDA \cite{Roth},
reported that 
by augmenting by one atom the core of their optimal 45-atom geometry 
consisting of a 38-atom outer shell and a 7-atom core,
a cluster of 46 Si atoms is obtained which has the same cohesive
energy, the same outer shell and the same number of 
``danging bonds'' as the original 
45-atom model.  In other words, adding one atom to
the 45-atom model gives a cluster of the same stability
(cohesive energy),
surface structure (outer shell) and electronic states
(number of dangling bonds).  Consequently, 
the reactivity of the 45 and 46 atom 
clusters obtained by these simulations
should be essentially identical, in direct contradiction to experiment
\cite{Smalley1,react1,react2,react3,Smalley2}.
The SRIG models presented here
give a natural explanation to this problem: since
the SRIG models (e.g. the 45-atom model with the lowest energy) 
correspond to a perfectly reconstructed outer shell, the addition 
or subtraction of one atom will drastically change the structure
because it cannot be accommodated as part of the outer shell
reconstruction or of the interior, and thus 
the reactivity of the cluster will be significantly increased. 
 
(3) The methodologies employed to search for possible cluster
structures in MD or SA simulations have limitations, 
especially when it comes to comparing geometries that
are very different in bonding arrangements. 
This applies to the empirical and semi-empirical methodologies, 
but also to a certain extend to the DFT-LDA methodologies.
It has been argued, for example, that electron correlation
effects are crucial in stabilizing the structure of certain Si clusters
\cite{Phillips13,Grossman-QMC}, 
while such effects are not explicitly included 
in the DFT-LDA simulations \cite{Roth,Mukher}.  
In the present work,
we have used the DFT-LDA approach 
{\em only for optimization of the structure} of models within
the proposed very strict geometric constraints, that is, as if the 
overall structure of the cluster were known.
It is well established that
this methodology gives reasonable agreement with experiment
for the detailed structural features (i.e. bond lengths
and bond angles) of molecules, when the overall geometry is known. 
The electronic structure corresponding to 
a known geometry is also reasonably well obtained.
There are however examples where this methodology
has proven inadequate in comparing the energetics of
very different structures for a cluster of a given size and
composition, for instance the cases where strong
correlation effects are present   
\cite{Phillips13,Grossman-QMC}. 
In this sense, it may be overly optimistic
to rely on DFT-LDA simulations in order to determine
the structure of Si clusters in the size range considered here,
without imposing any geometric constraints.
If this is true, 
the results of empirical or semi-empirical simulations
could be even less reliable, 
as far as unconstrained structure optimization of specific cluster sizes
is concerned.

(4) Even if a methodology that is very accurate were available,
the task of determining the optimal structure of clusters
in this size range is daunting.  The review of Jones and
Gunnarson \cite{Jones} presents persuasive arguments on the
intractability of finding the global energy minimum
for a structure with several tens of atoms through an unguided
search over configurational space.  In fact, it is very likely
that the structures with special properties (such as the magic numbers)
correspond to deep and narrow wells
in the multi-dimensional configurational space, which are
difficult to locate by unguided searches.  What we have
provided here are physically motivated models that
could potentially correspond to such deep and narrow wells
in the energy landscape.  The proper approach would then be
to use such models as starting points
for searches of the configurational space.  Even in this
case, extreme caution should be used in performing simulations,
since it is relatively easy to bring the structure outside
the well if the bottom has not been reached by thorough
relaxation.  
As an example, in the present work we have found that the cluster
geometry can be stuck in local energy minima and is prevented from
reaching the bottom of the energy well,  
simply due to the initial occupation 
of the electronic states.  For this reason, 
the number of electronic states used in a DFT-LDA simulation
must exceed 
the number of states required to accommodate the electrons, 
and the filling of the states near the 
Fermi level must be varied in order to achieve good
relaxation and to avoid getting trapped in local energy minima.
Once outside the minimum energy well, the simulation may end up
exploring irrelevant and uninteresting structures which
correspond to broad shallow basins in the energy landscape.

In conclusion, we have discussed a set of models for Si clusters
in the range 33 -- 61 atoms, which are physically motivated
and exhibit interesting patterns in their energy and 
electronic structure.  These models may be relevant to
understanding the exceptionally low chemical reactivity
of certain magic number sizes.  
The pattern of HOMO-LUMO gaps exhibited by these models
is consistent with the magic number clusters of low reactivity.
Finally, the models are compact clusters with little concave
surface area, and the lowest energy ones
tend to have small cross sections.     
It will be interesting to consider how these
models compare to the innovative experiments of Jarrold and coworkers
\cite{Jarrold-mobil},
on the gas phase mobility of Si clusters.

{\em Acknowledgement: }  This work was performed during the
Worskshop on Quantitative Methods in Materials Research at ITP,  
supported by 
the National Science Foundation 
under Grant No. PHY94-07194.

$^{\dagger}$ On leave of absence from Physics Department
and Division of Engineering and Applied Sciences, Harvard University.

\end{multicols}

\begin{table}
\label{TableI}
\begin{minipage}[t]{6.0in}
\caption{
Atomic positions in a.u. of representative atoms for the optimized structures
of clusters of sizes 33 -- 61.  Atoms labeled 1 are in the interior of
the cluster, the rest of the atoms are on the exterior and correspond to the 
atoms shown in Fig. 1 -- 8. 
}
\begin{tabular}{|l|r|r|r|r|r|r|r|r|r|r|}
    &  33        &  39A       &  39B       &  45A       &  45B       
    & 49A      & 49B         & 57       & 61A      & 61B               \\ \hline
$1(x)$ &$ 2.39471 $&$ 2.38887 $&$ 2.59713 $&$ 2.57556 $&$ 2.50590$ 
    &$ 2.48468 $&$-2.84402    $&$ 2.47912 $&$ 2.64227 $&$ 2.52290$\\
$1(y)$ &$ 2.39471 $&$ 2.38887 $&$ 2.59713 $&$ 2.57556 $&$ 2.50590$
    &$ 2.48468 $&$-2.84402    $&$ 2.47912 $&$ 2.64227 $&$ 2.52290$\\ 
$1(z)$ &$ 2.39471 $&$ 2.38887 $&$ 2.59713 $&$ 2.57556 $&$ 2.50590$
    &$ 2.48468 $&$-2.84402    $&$ 2.47912 $&$ 2.64227 $&$ 2.52290$\\ \hline
$2(x)$ &$ 5.21249 $&$ 5.16076 $&$ 2.47377 $&$ 2.99354 $&$-0.11844$
    &$ 0.85354 $&$ 1.15471    $&$ 3.09986 $&$ 3.16035 $&$ 1.74224$\\
$2(y)$ &$ 5.21249 $&$ 5.16076 $&$ 2.47377 $&$ 2.99354 $&$ 3.90217$
    &$ 4.39152 $&$ 4.53560    $&$ 3.09986 $&$ 3.16035 $&$ 5.59316$\\ 
$2(z)$ &$ 0.25158 $&$ 0.23991 $&$-5.51192 $&$ 7.03966 $&$ 5.80660$
    &$ 6.28003 $&$ 6.21782    $&$ 6.85742 $&$ 7.11649 $&$ 5.59316$\\ \hline
$3(x)$ &$ 1.52358 $&$ 1.65023 $&$ 3.39552 $&$ 1.18201 $&$ 1.73935$
    &$-2.03724 $&$-1.72237    $&$ 1.10146 $&$ 1.13596 $&$ 1.31967$\\ 
$3(y)$ &$ 1.52358 $&$ 1.50238 $&$ 3.39552 $&$ 4.13728 $&$ 7.67793$
    &$ 6.86991 $&$ 7.10602    $&$ 4.19919 $&$ 4.16677 $&$ 4.34534$\\
$3(z)$ &$-6.81708 $&$-7.11171 $&$ 6.87579 $&$-7.15181 $&$-4.25387$
    &$ 4.12951 $&$ 4.35574    $&$-7.33086 $&$-7.16307 $&$-8.39439$\\ \hline
$4(x)$ &$ 4.97057 $&$ 4.98998 $&$ 6.55082 $&$ 6.15942 $&$-0.46391$
    &$-0.85430 $&$-0.68414    $&$-1.76564 $&$ 5.29138 $&$ 5.41968$\\ 
$4(y)$ &$ 4.97057 $&$ 4.98998 $&$ 6.55082 $&$ 6.15942 $&$ 2.20720$
    &$ 2.05072 $&$ 2.06772    $&$ 1.76564 $&$ 5.29138 $&$ 5.41968$\\
$4(z)$ &$ 4.97057 $&$ 4.98998 $&$ 6.55082 $&$ 6.15942 $&$ 9.84219$
    &$ 9.58102 $&$ 9.29718    $&$11.08793 $&$-8.87922 $&$-9.17398$\\ \hline
$5(x)$ &$         $&$10.71204 $&$ 8.16900 $&$         $&$-5.11695$
    &$ 5.14767 $&$ 5.30305    $&$ 6.25738 $&$ 6.31190 $&$ 6.21939$\\ 
$5(y)$ &$         $&$ 0.00000 $&$ 0.00000 $&$         $&$-5.11695$
    &$ 5.14767 $&$ 5.30305    $&$ 6.25738 $&$ 6.31190 $&$ 6.21939$\\ 
$5(z)$ &$         $&$ 0.00000 $&$ 0.00000 $&$         $&$-5.11695$
    &$ 5.14767 $&$ 5.30305    $&$ 6.25738 $&$ 6.31190 $&$ 6.21939$\\ \hline
$6(x)$ &           &              &           &           &          
    &$-5.83946 $&$-5.61324    $&$         $&$-8.47946 $&$-8.50096$\\ 
$6(y)$ &           &              &           &           &           
    &$-5.83946 $&$-5.61324    $&$         $&$-8.47946 $&$-8.50096$\\ 
$6(z)$ &           &              &           &           &           
    &$-5.83946 $&$-5.61324    $&$         $&$-8.47946 $&$-8.50096$\\ 
\end{tabular}
\end{minipage}
\end{table}

\begin{table}
\label{TableII}
\begin{minipage}[t]{9.5cm}
\caption{
Structural, energetic and electronic features of SRIG models: 
Cross section $\Omega = \pi b_{min}^2$, 
ratio $R_c$ of concave triangles to total number of surface triangles,
cohesive energy per atom $E_c$, and   
HOMO-LUMO gap $E_g$,
[for the clusters that have no gap, 
the degeneracy and filling of the 
partially occupied level at the Fermi energy are given in brackets].
}
\begin{tabular}{|l|c|c|c|c|}
Cluster  & $\Omega$ (\AA$^2$) & $R_c$ & $E_c$ (eV/atom) & $E_g$ (eV) \\ \hline
33   & 52.2  & 0 / 960 ( 0 \%)     & 3.816  &  0.35  \\ 
39A  & 100.9 & 36 / 1034 (3.5 \%)  & 3.741  &  0.64  \\ 
39B  &  87.0 & 48 / 1156 (4.2 \%)  & 3.599  &  0.77  \\ 
45A  & 77.6  & 0 / 1544 ( 0 \%)    & 3.844  &  0.30  \\ 
45B  & 90.3  & 176 / 2896 (6.1 \%) & 3.807  & [2 (1/2)] \\ 
49A  & 86.0  & 36 / 1868  (1.9 \%) & 3.858  &  0.07  \\ 
49B  & 81.0  & 36 / 1868  (1.9 \%) & 3.855  & [3 (2/3)] \\ 
57   & 114.0 & 288 / 2672 (10.8 \%)& 3.712  &  0.04  \\ 
61A  & 141.4 & 240 / 2676 (9.0 \%) & 3.800  & [3 (2/3)] \\ 
61B  & 142.1 & 72 / 2128 (3.4 \%)  & 3.674  & [3 (1/3)] \\ 
\end{tabular}
\end{minipage}
\end{table}

\newpage
\begin{center}
FIGURE CAPTIONS
\end{center}

\bigskip

FIG. 1: Projected structure of the exterior 
atoms in the 33 model.  Three types of
atoms are shown, white apex atoms, hatched atoms and grey atoms.
The atoms of each kind are equivalent by symmetry.  The four
triangles outlined by dotted lines represent the four sides of
a tetrahedron, unfolded around the central triangle. The darker
solid lines represent covalent bonds between atoms.  The five 
tetrahedrally bonded interior atoms are not shown.  

\bigskip

FIG. 2: Same as in previous figure, for the model 39A.  There are four types
or atoms, white, hatched, grey and black. 

\bigskip

FIG. 3: Same as in previous figure, for the model 39B.

\bigskip

FIG. 4: Same as in previous figure, for the model 45A.  Three types of atoms
are shown, white, hatched and grey.

\bigskip

FIG. 5: Same as in previous figure, for the model 45B.  The chirality of
this model is evident in the zig-zag chains of atoms that straddle
the dotted lines.  This model also serves as the shell for the 49A and
49B models, in which the four apex
white atoms or the four face-center white atoms become equivalent to
adatoms on the (111) ($7 \times 7$) reconstruction, by the addition 
of a five interior atom core.  

\bigskip

FIG. 6: Same as in previous figure, for the 57 model. 

\bigskip

FIG. 7: Same as in previous figure, for the 61A model.  Five types of
atoms are shown, white apex, white face-center, hatched, grey and black.

\bigskip

FIG. 8: Same as in previous figure, for the 61B model.

\bigskip

FIG. 9: Density of states of the different SRIG models (shifted on
the vertical axis for clarity).  The height of
lines represents the degeneracy (1,2 or 3).  The 
vertical dashed line is the Fermi energy.

\end{document}